\documentclass[11pt,fleqn]{article}
\usepackage{graphicx}
\usepackage[a4paper,left=2cm,right=2cm]{geometry}
\usepackage{amsmath}
\usepackage{float}
\usepackage[latin1]{inputenc}
\usepackage{amsmath}
\usepackage{amsfonts}
\usepackage{amssymb}
\usepackage{float}
\floatstyle{boxed} 
\restylefloat{figure}

\newcommand{\be}{\begin{equation}}
\newcommand{\ee}{\end{equation}}
\newcommand{\bea}{\begin{eqnarray}}
\newcommand{\eea}{\end{eqnarray}}

\begin{document}
\title{$R(D^{(*)})$ and $\mathcal{B}r(B \rightarrow \tau\nu_{\tau})$ in a Flipped/Lepton-Specific 2HDM with anomalously
 enhanced charged Higgs coupling to $\tau$/b.}
\author{Lobsang Dhargyal \\\\\ Institute of Mathematical Sciences, Chennai 600113
, India.}
\date{26 November 2015}

\maketitle
\begin{abstract}
Babar, Belle and recently LHCb has reported an excess in the measurements of $R(D^{*})$, $R(D)$ and $\mathcal{B}r(B \rightarrow \tau \nu_{\tau} )$ than expected from SM, a possible signature of lepton flavor universality violating NP. In this work we analyze the phenomenological implications for these decay modes in a Flipped/Lepton-Specific 2HDM with anomalously enhanced Yukawa coupling of $H^{\pm}$ to $\tau$/b. When experimental and theoretical errors are added in quadrature, we conclude that this phenomenological extension of SM can give results in agreement within 1$\sigma$  deviation for the combination of $R(D^{(*)})$ and $\mathcal{B}r(B \rightarrow \tau \nu_{\tau})$ compare to about 4$\sigma$ deviation from SM from the latest combined[Babar,Belle,LHCb] experimental data for these observables.
\end{abstract}

\section{\large Introduction.}

It has been reported first by Babar\cite{Babar} and Belle\cite{Belle}, a possible hint of lepton flavor universality violating NP in $R(D^{(*)}) = \frac{Br(B \rightarrow D^{(*)}\tau\nu)}{Br(B \rightarrow D^{(*)}l\nu)}$. And recently LHCb\cite{LHCb} has also measured a deviation in $R(D^{*})$ from SM. The combined(Babar,Belle and LHCb) results gives\cite{Zoltan}
\be
\begin{split}
R(D)_{EXP} = 0.388 \pm 0.047\\
R(D^{*})_{EXP} = 0.321 \pm 0.021
\end{split}
\ee
Comparing these measurement with the SM predictions\cite{SM}
\be
\begin{split}
R(D)_{SM} = 0.297 \pm 0.017\\
R(D^{*})_{SM} = 0.252 \pm 0.003
\end{split}
\ee
there is a deviation of 1.8$\sigma$ for the R(D) and 3.3$\sigma$ for the $R(D^{*})$ and the combination corresponds close to a 3.8$\sigma$ deviation from SM prediction. It is further supported by measurement of Br($B \rightarrow \tau \nu$) by Babar\cite{Babar1} and Belle\cite{Belle1} and gives(from PDG average):
\be
Br_{EXP}(B \rightarrow \tau \nu) = (1.14 \pm 0.27)\times 10^{-4}
\ee
which is 1.3$\sigma$ above the SM prediction\cite{SM1}. Now if we add the errors in $R(D^{*})$ and $B(B \rightarrow \tau \nu)$ in quadrature then the three data combined deviates from SM by about 4$\sigma$.
\\
In this analysis we examine Higgs mediated flavor universality violation in a Flipped 2HDM with anomalous charged Higgs coupling to $\tau$ lepton. The outline of the paper is as follows: in section 2 we give a detailed analysis of Yukawa sector of Flipped 2HDM with anomalous charged Higgs coupling to $\tau$ lepton. In section 3 we give the decay rate and branching fraction formulas with form factor details. Followed by Results and Conclusions in the section 4.

\section{\large Flipped 2HDM with enhanced charged Higgs coupling to $\tau$ lepton.}

The effect of Physics beyond the SM in Taunic B decays has been studied extensively, particular in the context of Type-II 2HDM but Babar\cite{Babar} has ruled out Type-II 2HDM at 99.8 percent confidence level from the fits to the $R(D^{(*)})$ data for any values of $\frac{\tan\beta}{M_{\pm}}$. How ever Belle's measurements shows some comparability with Type-II 2HDM at about $\frac{\tan\beta}{M_{\pm}} = 0.5 GeV^{-1}$ but if we include the $B \rightarrow \tau \nu_{\tau}$ data also then it turn out contribution from such a large $\frac{\tan\beta}{M_{\pm}}$ actually pushes the $B \rightarrow \tau \nu_{\tau}$ well over 5$\sigma$ from the PDG world average. This is mainly due to the fact that since Type-I and Type-II 2HDMs interfere destructively with SM, in these type of 2HDMs only a large contribution from purely charged Higgs can have positive contribution from these models, but a large contribution from purely charged Higgs part that gives a moderate push to the $R(D^{(*)})$ pushes the $B \rightarrow \tau \nu_{\tau}$ beyond 5$\sigma$ from the world average very fast. So only a 2HDM with constructive interference with SM like Flipped or Lepton specific 2HDM may be able to fit all three of the B-physics data $R(D^{(*)})$ and $B \rightarrow \tau \nu_{\tau}$. But Flipped or Lepton specific 2HDM in itself does not fit because in these models the $\tan\beta$ dependence between $m_{b}$ and $m_{\tau}$ in the interference between SM and charged Higgs cancels out and so there is no corresponding increase in contribution from interference part as $\tan\beta$ increases. To fit all three of the data we need to modify these models so that there is some extra enhancement in interference part.

\subsection{\large Modifying Yukawa sector of Flipped 2HDM.}

In general 2HDM the Yukawa lagrangian is given as\cite{Branco}:
\be
\begin{split}
\mathcal{L}^{2HDM}_{Yukawa} = -\sum_{f=u,d,l}\frac{m_{f}}{\mathcal{V}_{0}}(\xi^{f}_{h}\bar{f}fh + \xi^{f}_{H}\bar{f}fH - i\xi^{f}_{A}\bar{f}\gamma_{5}fA)\\
                              -[\frac{\sqrt{2}V_{ud}}{\mathcal{V}_{0}}\bar{u}(m_{u}\xi^{u}_{A}P_{L} + m_{d}\xi^{d}_{A}P_{R})dH^{+} + \frac{\sqrt{2}\xi^{l}_{A}m_{\tau}}{\mathcal{V}_{0}}\bar{\nu_{L}}l_{R}H^{+} + H.c]
\end{split}
\ee
where $\xi$s depend on the type of 2HDM being used and $\mathcal{V}_{0} = 246$Gev is the vacuum expectation value of the Higgs. But since we require a constructive interference of SM and Charged Higgs contribution to fit all three of $R(D^{(*)})$ and $\mathcal{B}r(B \rightarrow \tau\nu_{\tau})$, only Lepton specific and Flipped 2HDM can achieve it. But contributions to the interference from just Lepton specific or Flipped 2HDM turn out to be too small to fit the three data simultaneously  since the $\tan\beta$ dependence between $m_{b}$ and $m_{\tau}$ cancels out in these models although they gives constructive interference with SM unlike Type-I and Type-II 2HDM. Some additional factor has to be introduce into these models so that it can fit the three data. One simplest and straight forward way to achieve this enhancement is if we require that $\tau$ lepton is screened from interacting with the full strength to the Higgs VEV $v_{2}$ and to all the neutral excitations from Higgs vacuum like the scalars h, $H^{0}$ and the pseudo-scalar $A^{0}$ in Flipped 2HDM. Since the Yukawa sector of the $\tau$ lepton breaks the $SU_{L}(2)$ symmetry the theory is not renormalizable in itself, so it has to be embed inside a larger model in which Flipped 2HDM comes out as a 400GeV-Few TeV scale effective theory\footnote{like Fermi's two current theory......}. The details of how such an anomalous interaction can be embed inside a larger model is out side the scope of the present work and will be assigned to a future work. Here we focus on the phenomenological consequences of such a screening on observed B decay anomalies. Now since $\tau$ lepton is screened from seeing the full depth of $v_{2}$, its Yukawa coupling must increase so that its mass which is proportional to $Y_{Yukawa}\times \frac{v_{2}}{\eta}$ now is the same as the observe mass $m_{\tau}$, which will effectively enhance the Yukawa coupling of $\tau$ lepton to the charged Higgs by a factor of $\eta$ while Yukawa coupling of $\tau$ to h, $H^{0}$ and $A^{0}$ remains same as in usual Flipped 2HDM as the $\eta$ factors in the neutral scalar Yukawa interaction cancels out with that of the $\eta$ factor coming from reduced higgs vacuum $\mathcal{V}_{0}$. For same analysis carried out in the Lepton Specific 2HDM, see the comments at the end of the paper. Then the $\xi^{f}_{A}$ factors in the charged Higgs interactions in this anomalous Flipped 2HDM with the screening factor $\eta$ is given by:
\be
\xi^{u}_{A} = \cot\beta$, $\xi^{d}_{A} = \tan\beta$ and $\xi^{e,\mu}_{A} = -\cot\beta$ and $\xi^{\tau}_{A} = -\eta\cot\beta.
\ee
Finally the charged Higgs lagrangian contributing to $b \rightarrow c l \nu $ in this model can be written in the most general form using Yukawa couplings derived from $\xi^{f}_{A}$s will be given as:
\be
\mathcal{L}^{Yukawa}_{H^{\pm}} = -[V_{cb} \bar{c}(g_{s} + g_{p}\gamma_{5})b H^{+} - \bar{\nu_{l}}(f_{s}^{l} + f_{p}^{l}\gamma_{5})l H^{+} + H.C ]
\ee
where\\

$g_{s} = \frac{(m_{b}\tan\beta + m_{c}\cot\beta)}{\sqrt{2}\mathcal{V}_{0}},  g_{p} = \frac{(m_{b}\tan\beta - m_{c}\cot\beta)}{\sqrt{2}\mathcal{V}_{0}},$
\be
f_{s}^{e,\mu} = f_{p}^{e,\mu} = \frac{m_{e,\mu}\cot\beta}{\sqrt{2}\mathcal{V}_{0}}$ and $f_{s}^{\tau} = f_{p}^{\tau} = \frac{m_{\tau}\cot\beta}{\sqrt{2}\mathcal{V}_{0}}\eta.
\ee
Note the relative negative sign between the hadronic current and leptonic current which will ensure constructive interference between SM and charged Higgs contributions. In this model, SM and charged Higgs interfere constructive unlike Type-I and Type-II 2HDM models where they interfere destructively. In this modified Flipped 2HDM, the charged Higgs coupling to $\tau$ is enhanced by a factor of $\eta$ than that of a simple Flipped 2HDM.\\
Then the effective langrangain given by charged Higgs exchange between the hadronic and leptonic currents for the $b \rightarrow cl\nu_{l}$ is given as:
\be
T_{H} = +\frac{V_{cb}}{M_{H}^{2}}[\bar{c}(g_{s} + g_{p}\gamma_{5})b\bar{l}(f^{l}_{s} - f^{l}_{p}\gamma_{5})\nu_{\bar{l}}].
\ee
And total effective langrangain of SM + charged Higgs for the $b \rightarrow cl\nu_{\bar{l}}$ is given as:
\be
T_{T} = \frac{G^{l}_{F}V_{cb}}{\sqrt{2}}[\bar{c}\gamma^{\mu}(1-\gamma_{5})b\bar{l}\gamma_{\mu}(1-\gamma_{5})\nu_{\bar{l}}]$ + $\frac{V_{cb}}{M_{H^{\pm}}^{2}}[\bar{c}(g_{s} + g_{p}\gamma_{5})b\bar{l}(f^{l}_{s} - f^{l}_{p}\gamma_{5})\nu_{\bar{l}}]
\ee
where $G_{F}$ is the Fermi coupling constant and $M_{H^{\pm}}$ is the mass of the charged Higgs.

\section{$R(D^{(*)})$ and $\mathcal{B}r(B \rightarrow \tau\nu_{\tau})$ in the model.}

\subsection{B $\rightarrow$ D $l$ $\nu_{l}$}

The standard parametrization of the hadronic matrix elements for the vector current is given as
\be
<D(P_{2})|\bar{c}\gamma^{\mu}(1-\gamma_{5})b|B(P_{1})>$ = $F_{1}(s)(P_{1} + P_{2})^{\mu} + \frac{m_{B}^{2} - m_{D}^{2}}{s}[F_{0}(s) - F_{1}(s)]q^{\mu}
\ee
and the standard parametrization of the hadronic matrix elements for the scalar current is given as
\be
<D(P_{2})|\bar{c}b|B(P_{1})>$ = $\frac{m_{B}^{2} - m_{D}^{2}}{m_{b} - m_{c}}F_{0}
\ee
where $F_{0}$ and $F_{1}$ are form factors.
\\
Then we have for the B $\rightarrow$ D $l$ $\nu_{l}$:
\be
\frac{d\Gamma_{SM}}{ds} = \frac{G_{F}^{2}|V_{cb}|^{2}}{96\pi^{3}m^{2}_{B}} \{  4m_{B}^{2}|\vec{P}_{D}|^{2}(1+\frac{m_{l}^{2}}{2s})|F_{1}|^{2} + [m_{B}^{4}(1-\frac{m_{D}^{2}}{m_{B}^{2}})^{2}\frac{3}{2}\frac{m_{l}^{2}}{s}]|F_{0}|^{2} \}(1-\frac{m_{l}^{2}}{s})^{2} |\vec{P}_{D}|
\ee
\be
\frac{d\Gamma_{MIX}}{ds} = +\frac{G_{F}}{\sqrt{2}}\frac{1}{M_{H}^{2}}\frac{m_{l}g_{s}|V_{cb}|^{2}}{32\pi^{3}}(f_{s}^{l} + f_{p}^{l})[(1-\frac{m_{D}^{2}}{m_{B}^{2}})]\frac{m_{B}^{2} - m_{D}^{2}}{m_{b} - m_{c}}F_{0}^{2}(1-\frac{m_{l}^{2}}{s})^{2}|\vec{P}_{D}|
\ee
\be
\frac{d\Gamma_{H}}{ds} = \frac{1}{M_{H}^{4}}\frac{g_{s}^{2}|V_{cb}|^{2}}{64\pi^{3}m_{B}^{2}}(\frac{m_{B}^{2} - m_{D}^{2}}{m_{b} - m_{c}})^{2}((f^{l}_{s})^{2} + (f_{p}^{l})^{2})F_{0}^{2}s(1 - \frac{m_{l}^{2}}{s})^{2}|\vec{P}_{D}|
\ee
where $|\vec{P}_{D}| = \frac{\sqrt{s^{2} + m_{B}^{4} + m_{D}^{4} - 2(sm_{B}^{2} + sm_{D}^{2} + m_{D}^{2}m_{B}^{2})}}{2m_{B}}$ is the momentum of the D in the B's rest frame and $g_{s}, g_{s}, f_{s}^{l}$ and $f_{p}^{l}$ are taken from Eqs(7). In terms of the Babar's parametrization\cite{Babar} we have
\be
F_{1} = \frac{\sqrt{m_{B}m_{D}}(m_{B} + m_{D})\sqrt{w^2 - 1}}{2m_{B}|\vec{P}_{D}|}V_{1}
\ee
\be
F_{0} = \frac{\sqrt{m_{B}m_{D}}(w + 1)}{m_{B} + m_{D}}S_{1}
\ee
where
\be
V_{1}(w) = V(1)[ 1 - 8\rho_{D}^{2}z(w) + (51\rho_{D}^{2} - 10)z(w)^{2} - (252\rho_{D}^{2} - 84)z(w)^{3} ]
\ee
\be
S_{1}(w) = V_{1}(w)\{1 + \Delta[ -0.019 + 0.04(w - 1) - 0.015(w - 1)^{2} ]\}
\ee
with  $\Delta = 1 \pm 1$ and
\be
w = \frac{m_{B}^{2} + m_{D}^{2} - s}{2m_{B}m_{D}}
\ee
\be
z(w) = \frac{(\sqrt{w + 1} - \sqrt{2})}{(\sqrt{w + 1} + \sqrt{2})}
\ee
\be
\rho_{D}^{2} = 1.186 \pm 0.055
\ee
and common normalization factor $V_{1}(1)$ cancels in the ratios.
\\
We take
\be
R(D) = \frac{\Gamma_{T}(B \rightarrow D \tau \nu_{\tau})}{\Gamma_{T}(B \rightarrow D l \nu_{l})}$ with $\Gamma_{T} = \Gamma_{SM} + \Gamma_{MIX} + \Gamma_{H^{\pm}}
\ee
where $l$ here refers to $\mu$ or e.

\subsection{B $\rightarrow$ $D^{*}$ $l$ $\nu_{l}$}

The hadronic matrix elements for the $B \rightarrow D^{*}$ is expressed in the terms of four QCD form factors $A_{1,2,3}(s)$ and V(s) as:
\be
<D^{*}(p_{D^{*}},\varepsilon^{*})|\bar{c}\gamma^{\mu}b|\bar{B_{p}}> = \frac{iV}{m_{B} + m_{D^{*}}}\varepsilon_{\mu\nu\alpha\beta}\varepsilon^{*\nu}p_{B}^{\alpha}p_{D^{*}}^{\beta}
\ee
\be
\begin{split}
<D^{*}(p_{D^{*}},\varepsilon^{*})|\bar{c}\gamma^{\mu}\gamma^{5}b|\bar{B_{p}}> = 2m_{D^{*}}A_{0}\frac{\varepsilon^{*}\cdot q}{s}q^{\mu} + (m_{B} + m_{D^{*}})A_{1}(\varepsilon^{*} - \frac{\varepsilon^{*}\cdot q}{s}q^{\mu})\\
-A_{2}\frac{\varepsilon^{*}\cdot q}{m_{B} + m_{D^{*}}}[(p_{B} - p_{D^{*}})^{\mu} - \frac{m_{B}^{2} - m_{D^{*}}^{2}}{s}q^{\mu}]
\end{split}
\ee
from which we get using equations of motion that scalar current vanishes while pseudo-scalar current is given as:
\be
<D^{*}(p_{D^{*}},\varepsilon^{*})|\bar{c}\gamma^{5}b|\bar{B_{p}}> = -\frac{2m_{D^{*}}}{\bar{m_{b}} + \bar{m_{c}}}A_{0}\varepsilon^{*}\cdot q
\ee
Then we have for the B $\rightarrow$ $D^{*}$ $l$ $\nu_{l}$:
\be
\frac{d\Gamma_{SM}}{ds} = \frac{G_{F}^{2}|V_{cb}|^{2}|\vec{P^{*}}|s}{96\pi^{3}m^{2}_{B}}(1-\frac{m_{\tau}^{2}}{s})^{2}[(|H_{+}|^{2} + |H_{-}|^{2} + |H_{0}|^{2})(1-\frac{m_{\tau}^{2}}{2s}) + \frac{3m_{\tau}^{2}}{2s}|H_{s}|^{2} ]
\ee
\be
\frac{d\Gamma_{MIX}}{ds} = +\frac{G_{F}}{\sqrt{2}}\frac{1}{M_{H}^{2}}\frac{m_{l}g_{p}|V_{cb}|^{2}}{8\pi^{3}}(f_{s}^{l} + f_{p}^{l})\frac{1}{m_{b} + m_{c}}A_{0}^{2}(1-\frac{m_{l}^{2}}{s})^{2}|\vec{P}_{D^{*}}|^{3}
\ee
\be
\frac{d\Gamma_{H}}{ds} = \frac{1}{M_{H}^{4}}\frac{g_{p}^{2}|V_{cb}|^{2}}{16\pi^{3}}\frac{1}{(m_{b} + m_{c})^{2}}((f_{s}^{l})^{2} + (f_{p}^{l})^{2})A_{0}^{2}(1 - \frac{m_{l}^{2}}{s} )^{2}s|\vec{P}_{D^{*}}|^{3}
\ee
where $|\vec{P}_{D^{*}}| = \frac{\sqrt{s^{2} + m_{B}^{4} + m_{D^{*}}^{4} - 2(sm_{B}^{2} + sm_{D^{*}}^{2} + m_{D^{*}}^{2}m_{B}^{2})}}{2m_{B}}$ is the momentun of $D^{*}$ in B's rest frame and $g_{s}, g_{s}, f_{s}^{l}$ and $f_{p}^{l}$ are taken from Eqs(7). In Babar's parametrization\cite{Babar} we have
\be
H_{\pm}(s) = (m_{B} + m_{D^{*}})A_{1}(s) \mp \frac{2m_{B}}{m_{B} + m_{D^{*}}}|\vec{P}_{D^{*}}|V(s)
\ee
\be
H_{0}(s) = \frac{-1}{2m_{D^{*}}\sqrt{s}}[ \frac{4m_{B}^{2}|\vec{P}_{D^{*}}|^{2}}{m_{B} + m_{D^{*}}}A_{2}(s) - (m_{B}^{2} - m_{D^{*}}^{2} - s)(m_{B} + m_{D^{*}})A_{1}(s)]
\ee
\be
H_{s}(s) = \frac{2m_{B}|\vec{P}_{D^{*}}|}{\sqrt{s}}A_{0}(s)
\ee
where
\be
A_{1}(w) = \frac{w + 1}{2}r_{D^{*}}h_{A_{1}}(w)
\ee
\be
A_{0}(w) = \frac{R_{0}(w)}{r_{D^{*}}}h_{A_{1}}(w)
\ee
\be
A_{2}(w) = \frac{R_{2}(w)}{r_{D^{*}}}h_{A_{1}}(w)
\ee
\be
V(w) = \frac{R_{1}(w)}{r_{D^{*}}}h_{A_{1}}(w)
\ee
with $w = \frac{m_{B}^{2} + m_{D^{*}}^{2} - s}{2m_{B}m_{D^{*}}}$ and $r_{D^{*}} = \frac{2\sqrt{m_{B}m_{D^{*}}}}{(m_{B} + m_{D^{*}})}$.
\\
The FF are given as:
\be
h_{A_{1}}(w) = h_{A_{1}}(1)[ 1 - 8\rho_{D^{*}}^{2}z(w) + (53\rho_{D^{*}}^{2} - 15)z(w)^{2} - (231\rho_{D^{*}}^{2} - 91)z(w)^{3} ]
\ee
\be
R_{0}(w) = R_{0}(1) - 0.11(w - 1) + 0.01(w - 1)^{2}
\ee
\be
R_{1}(w) = R_{1}(1) - 0.12(w - 1) + 0.05(w - 1)^{2}
\ee
\be
R_{2}(w) = R_{2}(1) + 0.11(w - 1) - 0.06(w - 1)^{2}
\ee
where $z(w) = \frac{(\sqrt{w + 1} - \sqrt{2})}{(\sqrt{w + 1} + \sqrt{2})}$ and $\rho_{D^{*}}^{2} = 1.207 \pm 0.028$, $R_{0}(1) = 1.14 \pm 0.07$, $R_{1}(1) = 1.401 \pm 0.033$ and $R_{2}(1) = 0.854 \pm 0.020$ and common normalization factor $h_{A_{1}}(1)$ cancels in the ratios and we take
\be
R(D^{*}) = \frac{\Gamma_{T}(B \rightarrow D^{*} \tau \nu_{\tau})}{\Gamma_{T}(B \rightarrow D^{*} l \nu_{l})}$ with $\Gamma_{T} = \Gamma_{SM} + \Gamma_{MIX} + \Gamma_{H^{\pm}}
\ee
where $l$ here refers to $\mu$ or e.

\subsection{$\mathcal{B}$r( $B^{-}$ $\rightarrow$ $\tau^{-}$ $\nu_{\tau}$)}

The total effective langrangain of SM + charged Higgs for the $B^{-} \rightarrow \tau^{-} \nu_{\tau}$ is given as:
\be
T_{T} = \frac{G_{F}V_{ub}}{\sqrt{2}}[\bar{u}\gamma^{\mu}(1-\gamma_{5})b\bar{\tau}\gamma_{\mu}(1-\gamma_{5})\nu] + \frac{V_{ub}}{M_{H}^{2}}[\bar{u}(g_{s}' + g_{p}'\gamma_{5})b\bar{\tau}(f_{s}^{\tau} - f_{p}^{\tau}\gamma_{5})\nu]
\ee
where $g_{s}' = \frac{(m_{b}\tan\beta + m_{u}\cot\beta)}{\sqrt{2}\mathcal{V}_{0}}$, $g_{p}' = \frac{(m_{b}\tan\beta - m_{u}\cot\beta)}{\sqrt{2}\mathcal{V}_{0}}$ and $f_{s}^{\tau}$, $f_{p}^{\tau}$ are taken from Eqs(7).
\\
The hadronic matrix elements of pseudo-vector current is given by
\be
<0|\bar{u}\gamma^{\mu}\gamma_{5}b|B> = iP_{B}^{\mu}f_{B}
\ee
and hadronic matrix elements of the pseudo-scalar current is given as
\be
<0|\bar{u}\gamma_{5}b|B> = if_{B0}
\ee
where $M_{H^{\pm}}$ is the mass of the charged Higgs and $f_{B0} = -\frac{m_{B}^{2}}{m_{b} + m_{u}}f_{B}$.
\\
Then we have for the B $\rightarrow$ $\tau$ $\nu_{\tau}$ :
\be
\mathcal{B}r_{SM}(B \rightarrow \tau \nu_{\tau}) = \frac{G_{F}^{2}|V_{ub}|^{2}m_{\tau}^{2}m_{B}}{8\pi}f_{B}^{2}(1 - \frac{m_{\tau}^{2}}{m_{B}^{2}})^{2} \tau_{B}
\ee
\be
\mathcal{B}r_{MIX}(B \rightarrow \tau \nu_{\tau}) = +\frac{G_{F}}{\sqrt{2}}\frac{1}{M_{H}^{2}}\frac{|V_{ub}|^{2}}{4\pi}g'_{p}f_{B}^{2}m_{\tau}\frac{m_{B}^{3}}{m_{b}+m_{u}}(f_{s}^{\tau} + f_{p}^{\tau})(1 - \frac{m_{\tau}^{2}}{m_{B}^{2}})^{2} \tau_{B}
\ee
\be
\mathcal{B}r_{H^{\pm}}(B \rightarrow \tau \nu_{\tau}) = \frac{1}{M_{H}^{4}}\frac{|V_{ub}|^{2}}{8\pi}f_{B}^{2}g_{'p}^{2}\frac{m_{B}^{5}}{(m_{b} + m_{u})^{2}}((f_{s}^{\tau})^{2} + (f_{p}^{\tau})^{2})(1 - \frac{m_{\tau}^{2}}{m_{B}^{2}})^{2} \tau_{B}
\ee
with
\be
\mathcal{B}r_{T} = \mathcal{B}r_{SM} + \mathcal{B}r_{MIX} + \mathcal{B}r_{H^{\pm}}
\ee
and $\tau_{B}$ is life time of B meson.

\section{Results and Conclusions.}

\subsection{Results:}

Although the $\eta$ is an independent parameter but if we require that $\eta = \tan^{2}\beta$, then it leads to very simple interpretation of the results. If we require $\eta = \tan^{2}\beta$, then the Yukawa interactions in the hadronic sector of our model has the same form as in the Type-II 2HDM but interaction in the leptonic sector of our model has the same form as in Flipped 2HDM except the $\tau$ lepton coupling to the charged Higgs, which has same form as in Type-II 2HDM but with opposite sign. Now there are only two parameters to fit i.e $\tan\beta$ and $M_{\pm}$, and a $\chi^{2}$ analysis with the $R(D^{(*)})$ and $\mathcal{B}$r( $B^{-}$ $\rightarrow$ $\tau^{-}$ $\nu_{\tau}$) as data points to find the best fits for the parameters $\tan\beta$ and $M_{\pm}$ gives $\chi^{2}_{min} = 10.597$ for $M_{\pm} >$ 400 GeV. This $M_{\pm} >$ 400 GeV constrain is from $B \rightarrow X_{s}\gamma$ and latest estimate on the lower bound of $M_{\pm}$ from $B \rightarrow X_{s}\gamma$ is given in \cite{Otto}. We have tabulated for three different values of the parameters that fits at same accuracy in the table below:\\
\\
\begin{table}[H]
\begin{center}
\begin{tabular}[b]{|c|c|c|c|c|c|} \hline
S.no & $\tan\beta$ & $M_{\pm}$ GeV & $R(D)_{Th}$ & $R(D^{*})_{Th}$ & $Br_{Th}(B \rightarrow \tau\nu)$ \\
\hline\hline
1 & 39.7 & 400 & 0.359 & 0.255 & 1.28$\times10^{-4}$ \\
\hline
2 & 69.5 & 700 & 0.359 & 0.255 & 1.28$\times10^{-4}$ \\
\hline
3 & 99.3 & 1000 & 0.359 & 0.255 & 1.28$\times10^{-4}$ \\
\hline
\end{tabular}
\end{center}
\caption{$\chi^{2}_{min} = 10.597$ and we have restricted the $\tan\beta$ in the range of $100 > \tan\beta > 1$.}
\end{table}

As the data in the table above and combine errors from experiments given in Eqs(1) and Eqs(3) shows, in the range 1 Tev $\ge M_{\pm} \ge$ 400 GeV and $100 > \tan\beta > 1$, we have:
\be
R(D)_{Th} = 0.359 \pm 0.15
\ee
\be
R(D^{*})_{Th} = 0.255 \pm 0.07
\ee
and
\be
Br_{Th}(B \rightarrow \tau\nu) = (1.28 \pm 0.88)\times10^{-4}
\ee
compared to the combined[Babar,Belle,LHCb]\cite{Zoltan} experimental values:\\
\be
R(D)_{EXP} = 0.388 \pm 0.047
\ee
\be
R(D^{*})_{EXP} = 0.321 \pm 0.021
\ee
and
\be
Br_{EXP}(B \rightarrow \tau \nu) = (1.14 \pm 0.27)\times 10^{-4}
\ee
\subsection{Conclusions:}

Babar, Belle and recently LHCb has reported an excess in the measurements of $R(D^{*})$, $R(D)$ and $\mathcal{B}r(B \rightarrow \tau \nu_{\tau} )$ than expected from SM, a possible signature of lepton flavor universality violating NP. In this work we have analyzed the implications for these decay modes in a Flipped 2HDM with enhanced Yukawa coupling of $H^{\pm}$ to $\tau$ lepton. By adding theoretical and experimental errors in quadrature from Eqs(48,49,50) and Eqs(51,52,53), we conclude that our phenomenological model can give results in agreement within 1$\sigma$  deviation for the combination of $R(D^{(*)})$ and $\mathcal{B}r(B \rightarrow \tau \nu_{\tau})$ compare to about 4$\sigma$ deviation from SM from the latest combined[Babar,Belle,LHCb] experimental data for these observables.\\
\\
$Comments:$ The same results can be achieved if b quark replaces the $\tau$ lepton in a Lepton Specific 2HDM. In that case Yukawa coupling of the leptonic sector will be same as Type-II 2HDM and Yukawa coupling of the quark sector will be same as Lepton Specific 2HDM\cite{Branco} except the b quark which will have the effective Yukawa coupling same as in Type-II with opposite sign. In that case charged Higgs mass may be allowed to be lower than 400 GeV and for a recent work on muon g-2 in Lepton Specific 2HDM and related bounds see \cite{Broggio}.

\newpage
{\large Acknowledgments: \large}  Author would like to thank Nita Sinha and Rahul Srivastava, Institute of Mathematical Sciences for helpful discussions and comments. Author would also like to thank Shrihari Gopalakrishna, Institute of Mathematical Sciences for helpful comments. This work is supported and funded by the Department of Atomic Energy of the Government of India and by the Government of Tamil Nadu.

\end{document}